\documentclass[prb,twocolumn,preprintnumbers,amsmath,amssymb,superscriptaddress]{revtex4}

\usepackage{graphicx}
\usepackage{dcolumn}
\usepackage{bm}
\usepackage{overpic}

\usepackage[normalem]{ulem} 
\usepackage{color} 

\newcommand{\bra}[1]{\ensuremath{\left\langle #1\right|}}
\newcommand{\ket}[1]{\ensuremath{\left|#1\right\rangle}}

\begin{document}

\title{Quantitative evaluation of defect-models in superconducting phase qubits}

\author{J.~H.~Cole}
\affiliation{Institut f\"ur Theoretische Festk\"orperphysik,
	Karlsruhe Institute of Technology, D-76128 Karlsruhe, Germany}
\affiliation{DFG-Center for Functional Nanostructures (CFN), D-76128 Karlsruhe, Germany}

\author{C.~M\"uller}
\affiliation{Institut f\"ur Theorie der Kondensierten Materie,
	Karlsruhe Institute of Technology, D-76128 Karlsruhe, Germany}
\affiliation{DFG-Center for Functional Nanostructures (CFN), D-76128 Karlsruhe, Germany}

\author{P.~Bushev}
\affiliation{Physikalisches Institut, Karlsruhe Institute of Technology,
D-76128 Karlsruhe, Germany}
\affiliation{DFG-Center for Functional Nanostructures (CFN), D-76128 Karlsruhe, Germany}

\author{G.~J.~Grabovskij}
\affiliation{Physikalisches Institut, Karlsruhe Institute of Technology,
D-76128 Karlsruhe, Germany}

\author{J.~Lisenfeld}
\affiliation{Physikalisches Institut, Karlsruhe Institute of Technology,
D-76128 Karlsruhe, Germany}

\author{A.~Lukashenko}
\affiliation{Physikalisches Institut, Karlsruhe Institute of Technology,
D-76128 Karlsruhe, Germany}

\author{A.~V.~Ustinov}
\affiliation{Physikalisches Institut, Karlsruhe Institute of Technology, D-76128 Karlsruhe, Germany}
\affiliation{DFG-Center for Functional Nanostructures (CFN), D-76128 Karlsruhe, Germany}

\author{A.~Shnirman}
\affiliation{Institut f\"ur Theorie der Kondensierten Materie,
	Karlsruhe Institute of Technology, D-76128 Karlsruhe, Germany}
\affiliation{DFG-Center for Functional Nanostructures (CFN), D-76128 Karlsruhe, Germany}

\date{\today}

\begin{abstract}
	We use high-precision spectroscopy and detailed theoretical modelling to determine the form of the coupling between a superconducting phase qubit 
	and a two-level defect.
	Fitting the experimental data with our theoretical model allows us to determine all relevant system parameters.  
	A strong qubit-defect coupling is observed, with a nearly vanishing longitudinal component.
	Using these estimates, we quantitatively compare several existing theoretical models for the microscopic origin of two-level defects.
\end{abstract}


\maketitle

A key limiting factor of superconducting quantum coherent devices is that they suffer from decoherence induced 
by their weak but non-negligible interaction with the environment~\cite{Shnirman:2002}.
The theoretical modelling of these interactions has greatly advanced our understanding of fundamental processes in the environment~\cite{Ithier:05} 
and led to improved designs for increased coherence times, 
e.g., by engineering `sweet-spots' or insensitivity to particular aspects of the environment~\cite{Vion:02, Koch:2007}.
Despite these advances, not all effects of the environment are understood. One such enigma is the appearence of pronounced anticrossings in the spectra of 
superconducting phase~\cite{Simmonds:04} and flux~\cite{Lupascu:09} qubits, which are indicative of a strong interaction with an additional quantum system.
It has be shown that these are coherent~\cite{Neeley:08} two-level, or at least strongly anharmonic~\cite{Bushev:2010}, defects, 
but their exact microscopic nature is still unclear.  

In several experiments~\cite{Neeley:08, Lupascu:09, Bushev:2010}, it has been observed that, for strongly coupled defects, the coupling term is transverse (involving pure qubit-defect energy exchange) with minimal longitudinal (phase shift inducing) component.  In this work, we perform a high precision comparison between experimental data and a general theoretical model to shed light on the exact form of the coupling operator between qubit and two-level defect. 
We obtain quantitative estimates of the longitudinal and transverse components and then compare our results to existing theoretical models for intrinsic two-level systems. 

We theoretically describe the system of qubit and two-level system (TLS) by the Hamiltonian 
\begin{equation}
	H = H_{q} + H_{\rm{TLS}} + H_{I} 
	\label{eq:H}
\end{equation}
where $H_{q}$ describes the qubit, $H_{\rm{TLS}}$ the TLS and $H_{I}$ the interaction between the two.
Our qubit is a flux biased phase qubit~\cite{Clarke:1988, Simmonds:04}, consisting of a superconducting ring interrupted by a Josephson junction and threaded by an external flux.
The qubit Hamiltonian is given by
\begin{equation}
	H_{q} = \frac{2 e^2}{C}\hat{q}^2 - E_J \cos \hat{\phi} + \frac{1}{2L} \left(\frac{\Phi_0}{2 \pi}\right)^2 \left(\hat{\phi} - \phi_{\rm{Ext}} \right)^2 \,,
	\label{eq:Hq}
\end{equation}
where $E_{J} = I_c \Phi_0/2 \pi$ is the Josephson energy of the circuit, $C$ is the qubit's capacitance, $L$ the inductance of the superconducting ring and
$\Phi_{0}$ is the superconducting flux quantum.
Eq.~(\ref{eq:Hq}) describes an anharmonic oscillator with dynamical variables given by the phase difference across the Josephson junction $\hat{\phi}$ 
and its conjugate momentum $\hat{q}$, corresponding to the number of cooper pairs tunneled across the junction, with $[\hat{q},\hat{\phi}]=i$.
The external flux $\phi_{Ext}$ is generated via a flux coil on chip. 
We assume a linear flux-current relation of the form $\phi_{Ext} = \alpha I_{\rm{bias}} + \beta$, 
with the fabrication dependent parameters $\alpha$ and $\beta$.  The TLS is described as a generic two-level system and we write its Hamiltonian in the eigenbasis
$H_{\rm{TLS}} = \frac{1}{2} \epsilon_{\rm{TLS}} \tau_{z}$, 
with the level splitting $\epsilon_{\rm{TLS}}$ and the Pauli-matrix $\tau_{z}$.

\begin{figure*}[t!]
      	\begin{overpic}[width=0.9\textwidth]{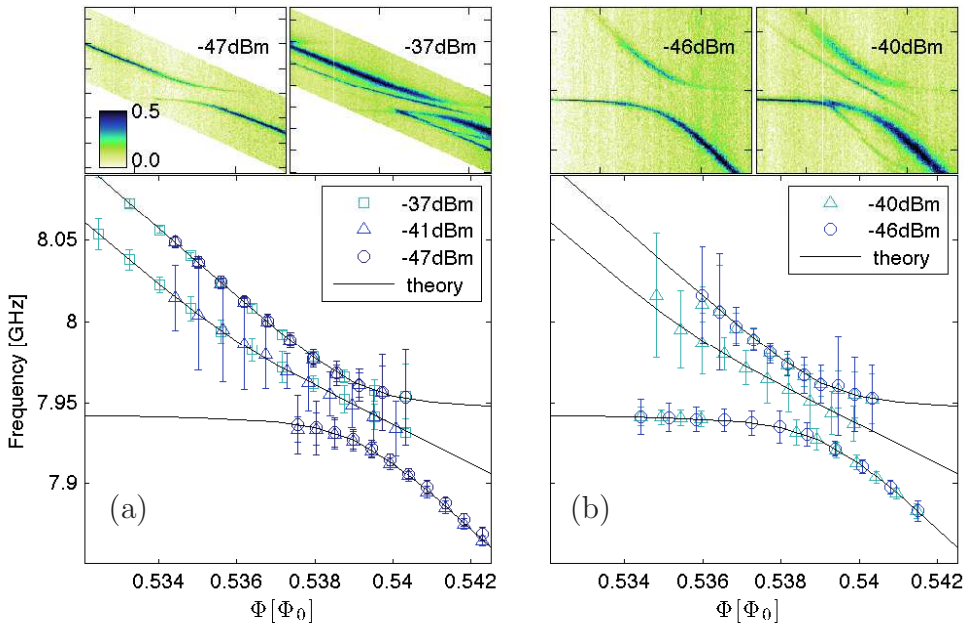}
		\put(15,10){\large (a)}
		\put(53,10){\large (b)}
	\end{overpic}
      	\caption{(color online) Peak positions obtained for (a) qubit spectroscopy and (b) swap spectroscopy.  For clarity, only 10\% of the dataset is shown.  
		The error bars give the 1-$\sigma$ confidence interval for the fitted peak positions.  
		The theoretical curves show the relevant transition frequencies for the coupled qubit-TLS system obtained 
		via fitting the extracted peak positions (see text).  
		Inserts show examples of the (normalized) escape probability as a function of excitation frequency and bias flux, from which the peak positions are extracted.
	}
      	\label{fig:spectra}
\end{figure*}

We consider three different coupling operators which may stem from fluctuations in the three terms of Eq.~(\ref{eq:Hq}), each of which corresponds to a different microscopic origin.  
The state of the TLS may modulate the magnetic flux $\phi_{Ext}$ threading the superconducting loop~\cite{Bluhm:2009, Sendelbach:2008} 
or the critical current $I_{c}$ of the Josephson junction~\cite{Constantin:2007, Ku:2005, deSousa:2009}, resulting in coupling to $\hat{\phi}$ or $\cos \hat{\phi}$, respectively.  Alternatively, the TLS may couple to the electric field of the junction $\vec{E}\propto \hat{q}$, which is consistent with the TLS being formed from a charge dipole~\cite{Martin:05, Martinis:2005}.  
These three situations are described by the following coupling Hamiltonians $H_{I}$,
\begin{eqnarray}
	H_{I}^{(\phi)} & = & v_\phi\: \hat{\phi} \: ( \cos{\theta_\phi}\: \tau_{x} + \sin{\theta_\phi}\: \tau_{z} )  
		\label{eq:HIphi} \\
	H_{I}^{(c)} & = & v_c\: \cos \hat{\phi} \: ( \cos{\theta_c}\: \tau_{x} + \sin{\theta_c}\: \tau_{z} ) 
		\label{eq:HIc} \\
	H_{I}^{(q)}  & = & v_q\: \hat{q} \: ( \cos{\theta_q}\: \tau_{x} + \sin{\theta_q}\: \tau_{z} )
		\label{eq:HIp} \,,
\end{eqnarray}
where $v_\phi$, $v_c$ or $v_q$ parameterize the strength of the coupling.  
The angles $\theta_\phi, \theta_c, \theta_q \in [0, \pi]$ denote the relative orientation of the TLS eigenbasis, the physical meaning of which depends on the particular microscopic model.  

In order to compare the various coupling models, 
we define the transverse $v_\perp$ and longitudinal $v_\parallel$ coupling in the qubit ($\ket{0}$, $\ket{1}$) 
basis as 
 \begin{eqnarray}
 	2 \, v_\perp  & = & v_{o} \cos \theta_{o} \left( \bra{1} \hat{o} \ket{0} + \bra{0} \hat{o} \ket{1} \right) \\
 	2 \, v_\parallel & = & v_{o} \sin \theta_{o} \left( \bra{1} \hat{o} \ket{1} - \bra{0} \hat{o} \ket{0} \right) ,
\end{eqnarray}
where $\hat{o} = \hat{q}$, $\hat{\phi}$ or $\cos \hat{\phi}$, the qubit component of the coupling term given be Eqs.~(\ref{eq:HIphi})-(\ref{eq:HIp}).  

To shed light on the nature of the interaction between qubit and two-level defect, we need to determine the values of $v$ and $\theta$.
To this end, we have performed a series of spectroscopy experiments of a superconducting phase qubit strongly coupled to a TLS, at varying microwave power~\cite{Bushev:2010}, see Fig.~\ref{fig:spectra}.  
Performing spectroscopy at both low- and high-power allows us to use a combination of single- and two-photon transitions 
to obtain spectral lines which are sensitive to the nature of the qubit-TLS coupling.
We also performed `swap-spectroscopy', where an additional swap between qubit and TLS is performed before readout, effectively measuring the state of the TLS.
We extract the frequencies of the various transitions in the coupled system by fitting each spectroscopic trace with Lorentzian functions. 

\begin{figure}[t!]
	\includegraphics[width=0.8\columnwidth]{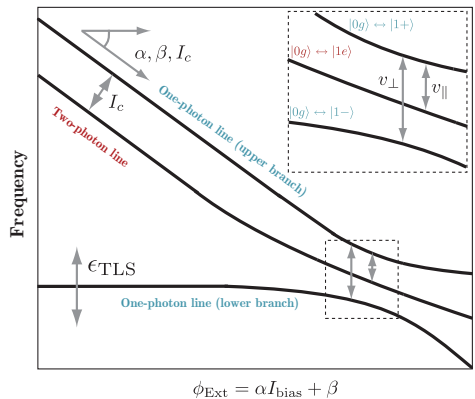}
	\caption{Anatomy of a qubit-TLS anti-crossing in the high power regime.  
		The overall slope of the spectral lines, their position and the spacing between one- and two-photon features allows us to calibrate the system, 
		even for several independent fitting parameters (see text).
		The separation and asymmetry of the lines, within the anti-crossing itself, allows us to estimate the transverse and longitudinal components of the coupling operator, respectively. 
	}
	\label{fig:Concept}
\end{figure}

Our theoretical model, Eq.~(\ref{eq:H}), can be described by a total of six independent parameters.
Three parameters describe the qubit circuit and its tuning via the external flux: the critical current $I_{c}$ of the qubits Josephson junction 
and the parameters $\alpha$ and $\beta$ describing the local generation of flux on chip and its coupling to the qubit loop.
The TLS is described by its level splitting $\epsilon_{\rm{TLS}}$ and the interaction between qubit and TLS via $v$ and $\theta$.
Figure~\ref{fig:Concept} shows an illustration of the spectrum of the model and the influences of the different parameters. 
Since their effects on the spectrum, as indicated by arrows in Fig.~\ref{fig:Concept}, are all largely independent, this allows us to perform a fit to all six parameters simultaneously.
For the circuit capacitance $C$ and inductance $L$ we take the design values of $C = 850$ fF and $L = 720$ pH. 
To account for fabrication variation, we repeated the fitting procedure with a $\pm5\%$ tolerance in both $L$ and $C$, 
resulting in no significant variation in the TLS parameter estimates (although $I_c$, $\alpha$ and $\beta$ vary accordingly).
It is important to note that, since we are limited to spectroscopic data, our results are only sensitive to purely transversal $\propto \sigma_{z} \tau_{z}$ 
and purely longitudinal $\propto \sigma_{x}\tau_{x}$ coupling terms. 

As an example, the estimated parameters for coupling to critical current according to Eq.~(\ref{eq:HIc}) are: 
level splitting $\epsilon_{\rm{TLS}} = 7944.38 \pm 0.08$ MHz with coupling strengths $v_{\perp} = 35.52 \pm 0.13$ MHz and $v_{\parallel} = 0.27 \pm 0.12$ MHz 
(uncertainties correspond to 1-$\sigma$ confidence intervals throughout).
We find the estimates obtained by fitting to each of the three coupling models are consistent with each other.
Repeating the fitting for an additional defect in the same chip with different level splitting $\epsilon_{\rm{TLS}}$ and coupling parameters $v$, 
$\theta$ produced qualitatively similar results, so we only consider one TLS in what follows.
Full details can be found in the supplementary material~\cite{Cole:supp:2010}.

We now discuss our results in light of several existing models describing the microscopic origin of such TLSs.  
Coupling to either magnetic flux or critical current generates both a transverse and longitudinal component.  
Using the ratio of these terms gives us an estimate for the orientation of $\tan \theta = 0.04 \pm 0.02$ for either coupling, 
placing strong constraints on critical current or magnetic flux coupling models.

If the state of the TLS modulates the value of the magnetic flux threading the superconducting loop, 
the observed coupling would result~\cite{Cole:supp:2010} from a magnetic flux contribution of $\delta \Phi_{Ext} = \delta \phi_{Ext} \Phi_{0}/(2\pi) \approx 250 \mu \Phi_{0}$. 
Assuming the fluctuations result from a spin in the surface of the superconducting loop of wire-thickness $\propto 1 \mu m$, 
such a modulation of the magnetic flux corresponds  to a magnetic moment approximately $10^{5}$ times that of an electron spin~\cite{Cole:supp:2010}. 

In the model of Ku et al.~\cite{Ku:2005}, a variation in critical current will couple to the qubit via the operator $\cos \hat{\phi}$. 
Using our estimates for $v_\perp$ and $v_\parallel$, we obtain~\cite{Cole:supp:2010} a critical current variation of $\delta I_{c} \approx 0.7$ nA, where $I_c = 984 \pm 2$ nA.
We can also use our estimates for $\tan\theta$ and $\epsilon_{\rm{TLS}}$ to calculate the TLS Hamiltonian in its \emph{physical} basis, 
where the two basis states correspond to different values of the qubits $I_{c}$. We obtain $H_{TLS} = 1/2 \: \epsilon_{0} \tau_{z} + 1/2 \: \Delta_{0} \tau_{x}$ 
with $\epsilon_{0} = 0.34 \pm 0.16$ GHz and $\Delta_{0} = 7.94 \pm 0.01$ GHz, giving two nearly degenerate states coupled by a large tunneling element. 
A similar calculation also holds for magnetic impurities.

Alternatively, the model of de Sousa et al.~\cite{deSousa:2009} assumes an impurity level in the junction which, via hybridisation with the Cooper-pairs in the superconductor, 
forms an Andreev bound state with energy inside the gap. Using this model~\cite{Cole:supp:2010} results in an impurity energy of $\epsilon_{d} \leq 150$ MHz 
and a variation in critical current of $\delta I_{c} \leq 1.5$ nA.  
Such an impurity energy that is close to the Fermi edge is a consequence of the small longitudinal coupling,  $\theta_c \approx 0$.

For a purely transverse coupling to the electric field, following Ref.~\onlinecite{Martinis:2005}, 
we can estimate the (aligned) dipole size (as a fraction of junction thickness $x$) for our TLS as $d/x = 0.08$.  
Since the momentum operator $\hat{q}$ has no diagonal component, 
this type of interaction would not lead to a longitudinal component $\propto \sigma_{z} \tau_{z}$ in the coupling operator.  Spectroscopy therefore provides no direct measure of the orientation $\theta_q$ of the charge-dipole.  Determining additional components of this form requires experiments which probe the dynamical properties of the system~\cite{Devitt:06}.

Although the data is compatible with a small longitudinal coupling (fitting to flux or critical current coupling), the resulting coupling strength $v_\parallel$ is comparable to the uncertainties and therefore we cannot rule out a pure charge-dipole.
In such a case, a small longitudinal coupling component may also stem from a variation in the junction potential along the lines of Ref.~\onlinecite{Constantin:2007}.  A linear combination of Eqs~\eqref{eq:HIphi}-\eqref{eq:HIp} is therefore also possible.

Using general theoretical models and high resolution spectroscopy, we have estimated the various coupling parameters between a superconducting phase qubit 
and a coherent two-level system within the qubit circuit.  Comparing with existing theoretical models, we obtained parameter estimates for various suggested sources of such defects.
In each case, the experimental data indicates a small or non-existent longitudinal coupling, relative to the transverse coupling term.  
These results allow us to place strong constraints on the parameters of the theoretical models and test their validity.

We would like to thank M. Ansmann and J. M. Martinis (UCSB) for providing us with the sample used in this work. 
This work was supported by the CFN of DFG, the EU projects EuroSQIP, MIDAS and SOLID, and the U.S. ARO under Contract No. W911NF-09-1-0336.

\bibliographystyle{apsrev}
\bibliography{lclimits}

\end{document}